\documentclass[twocolumn,showpacs,preprintnumbers,amsmath,amssymb]{revtex4}

\pdfoutput=1 

\usepackage{graphicx,subfigure}
\usepackage{dcolumn}
\usepackage{bm}
\usepackage{hyperref}
    
\usepackage{textcomp}
\usepackage{nicefrac}

\usepackage{mathptmx}
\usepackage{scrtime}
\usepackage{epstopdf}
\usepackage[normalem]{ulem} 
\usepackage{color}

\begin{document}

\title{Magnonic spin-wave modes in CoFeB antidot lattices}

\author{Henning Ulrichs}
\email{hulrich@gwdg.de}
\author{Benjamin Lenk}
\author{Markus M\"{u}nzenberg}
\affiliation{I. Institute of Physics, University of G\"{o}ttingen, Friedrich-Hund-Platz 1, 37077~G\"{o}ttingen, Germany}

\date{\today, \thistime}

\begin{abstract}
In this manuscript time-resolved magneto-optical Kerr effect experiments on structured CoFeB films are presented. The geometries considered are two dimensional square lattices of micrometer-sized antidots, fabricated by a focused ion beam. The spin-wave spectra of these magnonic crystals show a novel precessional mode, which can be related to a Bloch state at the zone boundary. Additionally, another magnetic mode of different nature appears, whose frequency displays no dependence on the externally applied magnetic field. These findings are interpreted as delocalized and localized modes, respectively.
\end{abstract}



\maketitle

Spin waves in confined magnetic structures have been investigated for a few decades and the idea to consider periodically modulated structures can be dated back to 1976~\cite{sykes}. Nowadays, magnetic metamaterials, i.e. materials with periodically varying magnetic properties in one, two or three dimensions, are called magnonic crystals.
The periodicity tailors the magnon band structure, eventually yielding band gaps and thus introducing the possibility to build (for example) spin-wave filters which have rather narrow transmission or rejection bands~\cite{Krawczyk,Chumak2008,Lee2009:mod:Py:stripes}.
Ultimately, the control over spin waves is necessary for the long-term purpose of building magnetism-based logic devices that use spin waves rather than electrons for information processing.
An important physical property of magnetic substructures is confinement caused by inhomogeneous demagnetizing fields~\cite{schloemann}. As a results, one finds localized modes in magnonic crystals, which are restricted to regions of strong internal field inhomogeneity~\cite{Neusser2008a,Jorzick2002:local:modes,Pechan}. These modes resemble atomic electronic states, which can extend throughout the whole crystal, if they get close enough to overlap. Regarding magnonic crystals, this situation is accomplished, if the inhomogeneity of the internal field connects the individual structural elements.
Considering micrometer scales and an experimental approach in the time domain, the emergence and experimental proof of spin waves is suppressed when the magnetic damping is too high. Hence, for any material composing magnonic crystals, a small magnetic damping is advantageous, as it is given for example in Permalloy ($\alpha=0.008$ \cite{PhysRevLett.101.237401}) or YIG ($\alpha=0.00006$ \cite{Goulon200738}). In the present paper $\mathrm{Co}_{20}\mathrm{Fe}_{60}\mathrm{B}_{20}$ has been investigated because of its particularly small damping constant of $\alpha=0.006$~\cite{Oogane}.
Also, CoFeB has the advantage of being amorphous, implying the absence of a magnetic anisotropy and grain boundaries and thus qualifying CoFeB as an ideal basic material for magnonic crystals.

The samples were prepared by magnetron sputtering CoFeB films with a thickness~$d$ of $50\,\mathrm{nm}$ onto a Si(100) substrate. A focussed ion beam technique (FIB) was used to create antidot structures with varying antidot radius~$r$ and lattice parameter~$a$ using a FEI Nova 600 NanoLab DualBeam. In all cases, the filling fraction was held constant at $f=\pi r^2 a^{-2}=0.12$, with the set of lattice parameters being $3.5$, $2.5$ and $1.5\,\text{\textmu m}$. After the FIB processing a top layer of $3\,\mathrm{nm}$ Ru was deposited by electron beam evaporation to prevent oxidation of the ferromagnetic layer.

The experiments were carried out with an all-optical pump probe setup, which exploits in a time-resolved manner the magneto-optical Kerr effect (TRMOKE). High intensity pump pulses impinge on the sample surface with a fluence of $57\,\mathrm{mJ\,cm^{-2}}$ and the successive temperature rise triggers the magnetic precession via an effective field pulse~\cite{Kampen2002,liu:JAP:2007}.
The excited high-k magnons decay within a few ps into spin waves with typical frequencies of a few GHz~\cite{marija07} and these magnetic oscillations are detected with probe pulses, whose Kerr rotation upon reflection from the sample is proportional to the (dynamic) magnetization. With this technique, angular amplitudes of the excited spin waves of several degrees, depending on the pump fluence, can be reached.
Note that the overall size of the structured regions was $(150\,\mathrm{\text{\textmu} m})^2$. While this cannot be considered an infinitely extended crystal, it is still large enough for our experimental approach, as the pump (probe) pulses are focussed to a Gaussian width of $60\,\mathrm{\text{\textmu}m}$ ($18\,\mathrm{\text{\textmu}m}$), respectively. The setup has previously been described in more detail elsewhere~\cite{walowski2008}.

\begin{figure}
	\centering
    \includegraphics[width=0.98\linewidth]{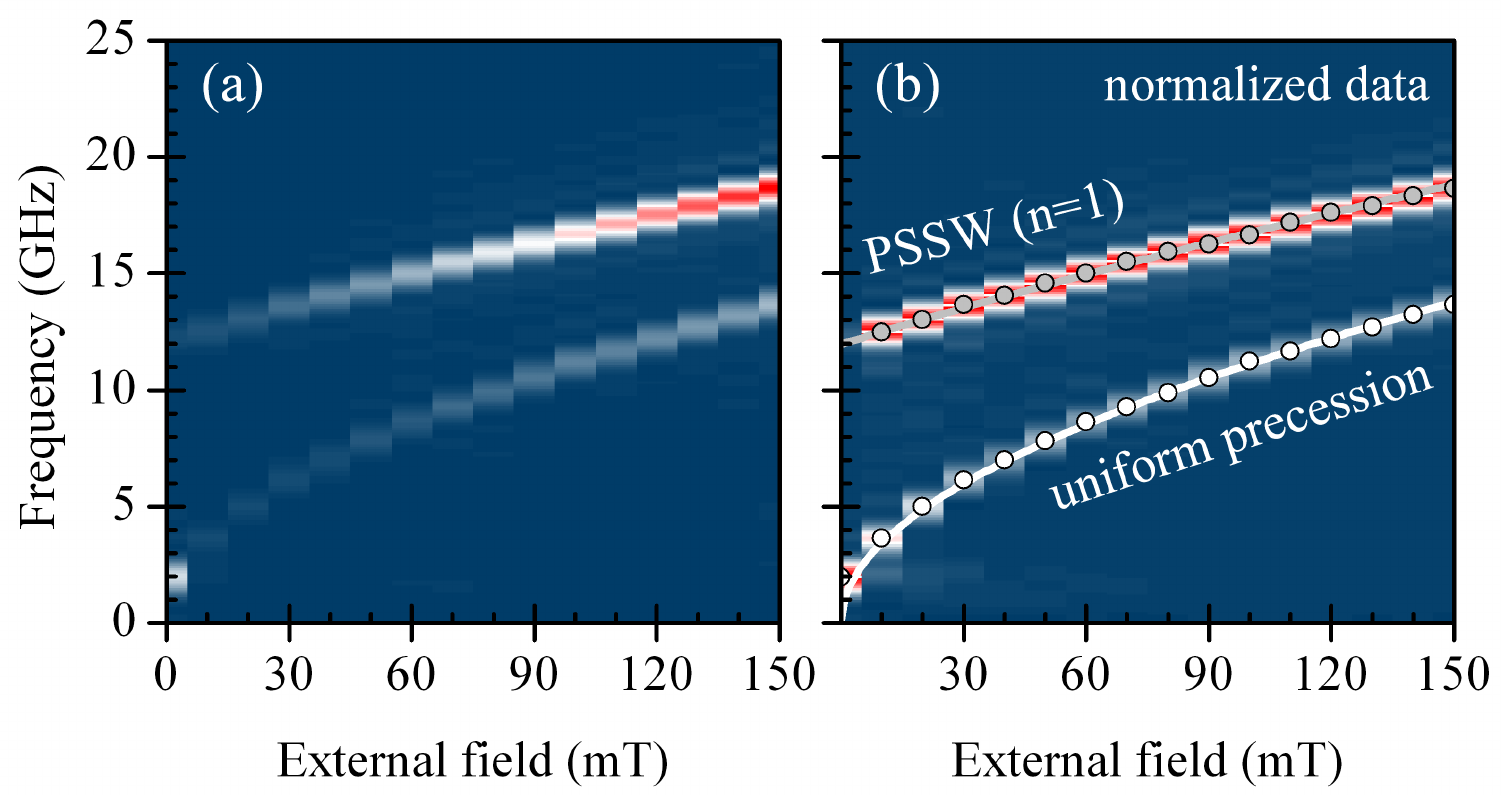}
    \caption{(color online). Magnetization dynamics on a continuous CoFeB film of thickness $d=50\,\mathrm{nm}$. (a) Fourier power spectrum as calculated from the time-resolved MOKE data~\cite{walowski2008}. (b) After normalization the frequency branches can be attributed to the uniform precession (white) as well as perpendicular standing spin waves (gray). Points are the peak positions, lines represent the fitted dispersion ~\eqref{gl:PSSW} for $n=0$ and $n=1$, respectively.}
	\label{fig:conti}
\end{figure}

Magnetization dynamics were measured covering a time interval of up to $1\,\mathrm{ns}$ after the arrival of the pump pulse. A quantitative analysis was carried out by Fourier transformation. To minimize artefacts stemming from the (discrete) transformation, a double exponential background originating from incoherent magnons and phonons was subtracted from the raw data prior to calculation of the oscillations' FFT power. A series of TRMOKE spectra for systematically varied external fields can be processed this way and the resulting Fourier power spectra can be plotted in a color code in the frequency-field plane. In the present paper only readily analyzed data sets are presented, a more detailed example of the analysis is also given in Ref.~\cite{walowski2008}.
From \hyperref[fig:conti]{\autoref{fig:conti}(a)} it is apparent, that for small external fields the spin-wave amplitude strongly diminishes, therefore normalization of each power spectrum to its respective maximum provides a better insight as shown in \hyperref[fig:conti]{\autoref{fig:conti}(b)}. The evolution of the observed magnetic modes with the applied field can then be interpreted with respect to the origin of the different modes.

The spin-wave theory for continuous films \cite{Kalinikos} provides the theoretical framework for the classification of the different modes. For an in-plane propagation direction which is perpendicular to the applied field, spin waves are called Damon-Eshbach (DE) modes \cite{Damon}. Neglecting the exchange contribution, the dispersion relation valid for small $k$ reads:
\begin{eqnarray}
\omega_{DE}^2=\omega_H\left(\omega_H+\omega_M\right)+\frac{\omega_M^2}{4}\left(1-e^{-2kd} \right).\label{gl:DE}
\end{eqnarray}
Within the same approximation the expression for an angle of $45^\circ$ between $k$ and $H$ reads:
\begin{multline}
\omega_{45^\circ}^2 =\omega_H^2+\frac{\omega_H\omega_M}{2}\left(1+\frac{1}{kd}\left[1-e^{-kd}\right] \right)+\\
\frac{\omega_M^2}{2kd}\left[1-e^{-kd}\right]\left(1-\frac{1}{kd}\left[1-e^{-kd}\right]\right).\label{gl:disp45}
\end{multline}
If one allows for propagation in the direction normal to the film plane, one ends up with exchange dominated perpendicularly standing spin waves (PSSW):
\begin{align}
\omega_{PSSW}^2=\left(\omega_H+\omega_A n^2\right)\left(\omega_H+\omega_A n^2+\omega_M\right).\label{gl:PSSW}
\end{align}
For $n=0$ Eq.~\eqref{gl:DE} describes the Kittel mode, which is characterized by a uniform precession of all spins. The abbreviations in Eqs.~\eqref{gl:DE}, \eqref{gl:disp45} and~\eqref{gl:PSSW} are related to the external field $H_\mathrm{ext}$ via $\omega_H=\gamma\mu_0 H_\mathrm{ext}\cos 30^\circ$, to the saturation magnetization $M_\mathrm{S}$ via $\omega_M=\gamma\mu_0 M_\mathrm{S}$ and to the exchange stiffness constant $A$ via $\omega_A=\gamma\frac{2A}{M_\mathrm{S}}\left(\frac{\pi}{d}\right)^2$, where $\gamma=g\mu_B \hbar^{-1}$ is determined by the gyromagnetic ratio. The trigonometric factor in $\omega_H$ accounts for the experimental necessity of a certain canting of the external field with respect to the sample's plane.

The spectrum of the unstructured film shows two different modes: the Kittel mode ($n=0$) and the first order standing spin wave ($n=1$). In combination with a SQUID measurement of $M_\mathrm{S}$ the following set of material parameters could be derived. The g-factor $g=2.04(1)$ as compared to $2.09$~\cite{Paluskar} is a little smaller than expected, whereas the saturation magnetization of $\mu_0M_\mathrm{S}=1.8(2)\,\text{T}$  and the low damping constant $\alpha=0.006(1)$ match values found by other groups~\cite{handley,Oogane}. The exchange stiffness constant $A=2.0(1)\times 10^{-11}\,\mathrm{J\, m^{-1}}$ is smaller than the commonly used value $2.8(2)\times 10^{-11}\,\mathrm{J\, m^{-1}}$ \cite{Bilzer}, which belongs to the different composition Co$_{72}$Fe$_{18}$B$_{10}$.
\begin{figure}
	\centering
    \includegraphics[width=0.98\linewidth]{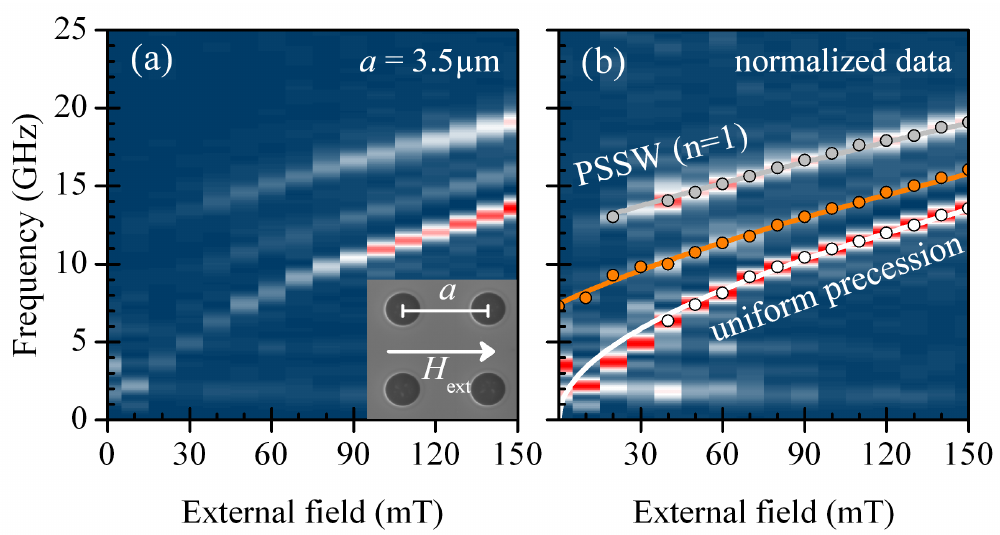}
    \caption{(color online).  Not normalized (a) and normalized (b) power spectra of the magnetization dynamics on an array of antidots. The inset in (a) shows a SEM picture of an antidot structure and schematic of the field geometry. As compared to \autoref{fig:conti}, an additional magnetic mode is observed (orange), which can be identified as a Damon-Eshbach wave, owning the wavelength given by the periodicity of the antidots.}
	\label{fig:rem-00deg}
\end{figure}
\begin{figure*}
	\centering
    \includegraphics[width=0.98\linewidth]{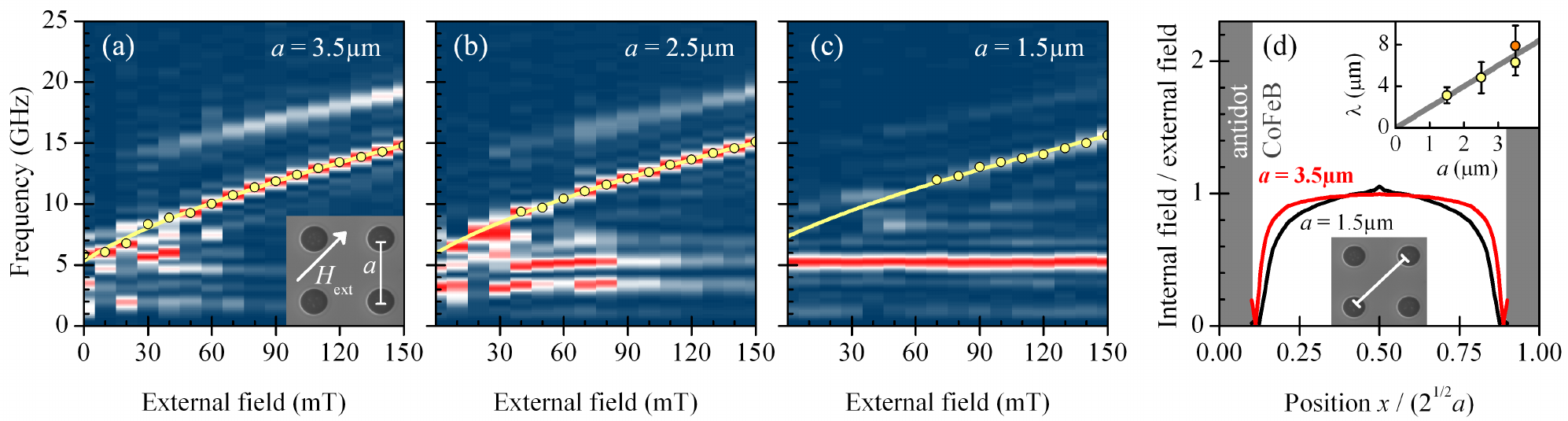}
    \caption{(color online). (a) to (c) Variation of antidot separation $a$ for an external field orientation of $\varphi=45^\circ$. Solid yellow lines show fits of Eq. \eqref{gl:disp45} to field dependent modes (yellow dots). (d) Numerical computation of the internal field $H_\mathrm{int}=H_\mathrm{ext}+H_\mathrm{dem}$ along the white section shown in the SEM picture. The external field $\mu_0H_\mathrm{ext}=140\,\mathrm{mT}$ is oriented parallel to the section; the inset shows the wavelength as a function of the antidots distance $a$, stemming from the fits depicted in (a) to (c); orange dot from the fit of Eq. \eqref{gl:DE}, shown in Fig.~\ref{fig:rem-00deg}(b).}
	\label{fig:psi45}
\end{figure*}
The Fourier power spectra in Fig.~\ref{fig:rem-00deg} show, that in the structured region a new mode appears, whose frequency is between the Kittel mode and the first order PSSW. The excitation mechanism does not select any propagation direction, wavelength or frequency. The population of the possible spin-wave modes is rather governed by the density of states arising from the magnonic band structure.
This density is high, wherever the dispersion is flat. According to this argument, a k-vector at the boundary of the first magnonic Brillouin zone can be expected, which means $k={\pi}{a^{-1}}$.
Assuming a propagation direction perpendicular to the applied field, a fit of Eq.~\eqref{gl:DE} to the newly appeared mode yields approximately this value (see Fig.~\ref{fig:rem-00deg} and Fig.~\ref{fig:psi45}(d)).
Higher order DE modes ($k=n\pi a^{-1}$) are less propable to detect due to damping. From the observation of the frequency alone one cannot judge, whether the propagation direction is $90^\circ$ or $270^\circ$ with respect to the external field, because the spin-wave theory \cite{Kalinikos, Damon} provides both possibilities.
The particular property of DE modes of being localized at one side of the film can be neglected in the present case, since $\lambda \gg d$. In fact, it seems most probable, that spin waves travelling in both directions were excited. Then, the resulting superposition describes a 1d standing wave pattern, with vanishing group velocity. \cite{footnote1}

Rotating the sample in-plane to $\varphi=45^\circ$ alters the spectrum, so that again only two modes can be identified (see Fig. \ref{fig:psi45}). At first sight the lower mode seems to be the Kittel mode as observed in the spectrum of the continuous film. A closer look reveals that this mode is shifted towards higher frequencies.
The new mode for $\varphi=0^\circ$ has a k-vector which points along the next nearest neighbour direction. For $\varphi=45^\circ$ this crystallographic direction has a different orientation relative to the external field: now this mode should be described by Eq.~\eqref{gl:disp45}.
Fitting Eq.~\eqref{gl:disp45} to the dominant peaks in Fig.~\ref{fig:psi45}(a) (yellow line) indeed yields again $k={\pi}{a^{-1}}$. The measurements and analysis were carried out for all three different lattice constants. The experimental power spectra always show a mode, which is in frequency slightly above the Kittel mode and the fit confirmed in each case that $k={\pi}{a^{-1}}$ (see Fig.~\ref{fig:psi45}(d)). For $\varphi=45^\circ$ the observed quantized modes are much stronger than the quantized DE mode seen in the spectrum for $\varphi=0^\circ$.
A simple reason could be, that the frequency is fourfold degenerate: spin waves with $k=\pi a^{-1}$ propagating in directions of $45^\circ$, $135^\circ$, $225^\circ$ and $315^\circ$ relative to the external field have the same frequency and can be superposed to a 2d standing wave pattern.
Although the discussed modes may possess stationary wave patterns, they originate from the delocalized modes described by the spin-wave theory of continuous films \cite{Kalinikos}. Owning their wavelength from the structural periodicity, these modes can be denoted as delocalized Bloch modes.

It can be argued, that the detection of spin waves, which have a finite wavelength, with the given setup should at least be strongly suppressed by the fact, that the probe laser pulse averages over a certain area.
Thus opposing phases contribute to the net signal. If they do so equally, the different contributions should indeed cancel each other. But since the probe pulses display a gaussian intensity profile in the lateral directions, opposing phase contributions in different unit cells of the antidot lattice do not cancel totally.

In addition to these findings, a second set of modes appears with decreasing lattice constant, whose frequencies only weakly depend on the strength of the external field (see Fig.~\ref{fig:psi45}). A simple ansatz to explain the field independence is to assume, that the observed oscillations belong to phononic excitations, such as surface acoustic waves. In the time-resolved reflectivity signal, indeed a weak oscillation is observed. The associated frequency of roughly $0.5\,\mathrm{GHz}$ does not compare to the frequencies which appear in the time-resolved MOKE spectra. Although the discussed modes do not significantly depend in frequency on the strength of the applied field, changing the relative orientation of $H_{\mathrm{ext}}$ alters this frequency, or introduces a second or third mode with a fixed but different frequency. Also, resonance effects can be seen in not normalized spectra, when the frequencies of the different sets of modes intersect. This clearly rules out a phononic origin.
The following consideration relates the appearance of the new mode to the internal field and suggests that the mode is a localized spin wave, existing close to each antidot's borders and  oscillating in phase throughout the whole crystal.
The geometric filling fraction $f=0.12$ remains the same for the different lattice constants, whereas the fraction of the regions with a not saturated, inhomogeneous internal field grows with a shrinking lattice constant. A numerical computation of the internal field, using the software package Nmag \cite{Nmag}, illustrates this behavior.
The internal field is given in Fig.~\ref{fig:psi45}(d). We expect that localized spin-wave modes should therefore become more and more dominant.
For $a=3.5\,\text{\textmu m}$ the new modes can almost not be distinguished from noise and the modes known from the continuous film and the delocalized mode dominate the spectrum. For $a=1.5\,\text{\textmu m}$, this situation is completely reversed, as can be seen in Fig.~\ref{fig:psi45}(c). Now regions with a saturated internal field $H_\mathrm{int}=H_\mathrm{ext}$ only become relevant enough in size to significantly allow for delocalized spin waves, when the external field exceeds a certain threshold value, which increases with a decreasing lattice constant. \cite{footnote2} This shows, that apart from the absolute size of the structures, the external field can be used to tune the spin-wave spectrum of the same magnonic crystal between a metallic and a more atomic behavior, with delocalized and localized spin waves dominating, respectively.

In summary an all-optical pump-probe technique was used to investigate spin-wave spectra of structured CoFeB films. The spectra of this kind of 2d magnonic crystal show modes with a k-vector $k=\pi a^{-1}$ induced by the period~$a$ of the magnonic crystal. Reducing the lattice constant from $3.5$ to $1.5\,\text{\textmu m}$, new modes apear which dominate for small~$a$. Our explanation suggests, that these modes are localized in regions of strongly inhomogeneous internal field. Research was funded by DFG priority program 1133.

\begin{thebibliography}{23}
\bibitem[{\citenamefont{footnote}()}]{footnote1}
\bibinfo{author}\bibnamefont{In comparison with the unstructured film the Kittel mode seems much stronger in power inside the structure. Note that a quantized Backward-Volume mode ($\mathbf k\| \mathbf H$) with $k=\pi a^{-1}$ has a frequency, which is so close to the Kittelmode, that the frequency resolution in the experiment ($0.5\,$GHz) is too large to distinguish the modes. This consideration indicates, that quantized Backward-Volume modes may also contribute to the oscillations denoted simply as uniform precession in Fig.~\ref{fig:rem-00deg}.}.
\bibitem[{\citenamefont{footnote}()}]{footnote2}
\bibinfo{author}\bibnamefont{This can also be seen in numerical simulations of the internal field. To avoid overcrowding, a field variation is not included in Fig.~\ref{fig:psi45}(d). Considering the fraction of the inhomogeneous internal field, the effect of reducing the lattice constant is qualitatively the same as for decreasing the external field: in both cases the fractional area of non-saturated $H_\mathrm{int}$ grows.}.
\end{thebibliography}

\end{document}